\documentclass[letterpaper]{article} %
\usepackage{aaai25}  %
\usepackage{times}  %
\usepackage{helvet}  %
\usepackage{courier}  %
\usepackage[hyphens]{url}  %
\usepackage{graphicx} %
\usepackage{natbib}  %
\usepackage{caption} %
\usepackage{algorithm}
\usepackage{algorithmic}

\usepackage{newfloat}
\usepackage{listings}
\DeclareCaptionStyle{ruled}{labelfont=normalfont,labelsep=colon,strut=off} %
\floatstyle{ruled}
\newfloat{listing}{tb}{lst}{}
\floatname{listing}{Listing}
\usepackage{booktabs}
\usepackage{multirow}

\title{Catastrophic Liability: Managing Systemic Risks in Frontier AI Development}
\author {
    Aidan Kierans\textsuperscript{\rm 1},
    Kaley Rittichier\textsuperscript{\rm 2},
    Utku Sonsayar\textsuperscript{\rm 2},
    Avijit Ghosh\textsuperscript{\rm 3,1}
}
\affiliations {
    \textsuperscript{\rm 1}School of Computing, University of Connecticut\\
    \textsuperscript{\rm 2}Philosophy Department, University of Connecticut\\
    \textsuperscript{\rm 3}Hugging Face\\
    aidan.kierans@uconn.edu, kaley.rittichier@uconn.edu, utku.sonsayar@uconn.edu, avijit.g@uconn.edu
}

\begin{document}
\nocopyright

\maketitle

\begin{abstract}
As artificial intelligence systems grow more capable and autonomous, frontier AI development poses potential systemic risks that could affect society at a massive scale. Current practices at many AI labs developing these systems lack sufficient transparency around safety measures, testing procedures, and governance structures. This opacity makes it challenging to verify safety claims or establish appropriate liability when harm occurs. Drawing on liability frameworks from nuclear energy, aviation software, cybersecurity, and healthcare, we propose a comprehensive approach to safety documentation and accountability in frontier AI development.
\end{abstract}

\section{Introduction}
As the capabilities of artificial intelligence (AI) systems rapidly advance, it is increasingly clear that frontier AI development has the potential to pose systemic risks to humanity if it is not done safely and responsibly. Policymakers, researchers, and industry leaders recognize the need to prioritize establishing robust frameworks for AI safety and governance~\cite{cais2023ai_risk}.

The risks from AI emerge during development, not just adoption; if an advanced AI system escapes control to pursue its own goals, or is stolen and misused to accelerate chemical, biological, or nuclear weapons development, the damages could be catastrophic~\cite{slattery2024airiskrepositorycomprehensive}. Yet the practices of companies developing frontier AI systems are often highly opaque, with little transparency into their safety practices, testing procedures, or governance structures for managing risks. This leaves the critical task of verifying the safety of these immensely powerful AI systems largely in the hands of the companies themselves or third-party evaluators contracted to test the final products.

The resulting lack of oversight is deeply concerning given the stakes involved. Furthermore, competition dynamics incentivize AI companies to race ahead in capabilities at the expense of thorough safety practices~\cite{walter_rapid_2024,bengio2024managing}. Thus, policymakers and researchers have an interest in assessing whether self-reported safety claims truly mitigate key risks, but the opaque nature of AI development makes this difficult or impossible.

\paragraph{Thesis and approach.} We argue that greater transparency and documentation of risk management practices can simultaneously increase safety and reduce costs for frontier AI development. Each of the major stakeholders involved—policymakers, standards bodies, and industry—can advance their own interests while supporting each other by describing and demonstrating responsibility. This mutually beneficial dynamic emerges from three key contributions:

\textbf{(1) We establish that voluntary AI standards already carry legal force through existing tort law.} U.S. courts determine negligence based on "reasonable practice" rather than merely common practice, giving frameworks like NIST's AI Risk Management Framework potential legal weight even without mandatory regulation. This challenges the assumption that voluntary commitments lack enforcement mechanisms.

\textbf{(2) We demonstrate that frontier AI already poses systemic risks requiring expanded duties of care.} Unlike high-risk AI applications that create dangers through specific deployments, general-purpose models with systemic risk can pose threats through their underlying capabilities—regardless of intended use. The EU AI Act recognizes this distinction by imposing obligations on frontier model providers based on computational thresholds and capability assessments, not just deployment contexts.

\textbf{(3) We show how comprehensive documentation creates competitive advantages while improving safety.} Through case studies from nuclear energy, aviation, healthcare, and cybersecurity, we demonstrate that transparency reduces liability exposure for responsible developers while enabling meaningful oversight. Current opacity leaves all parties worse off.

\begin{figure*}[ht]
\centering
\includegraphics[width=2\columnwidth]{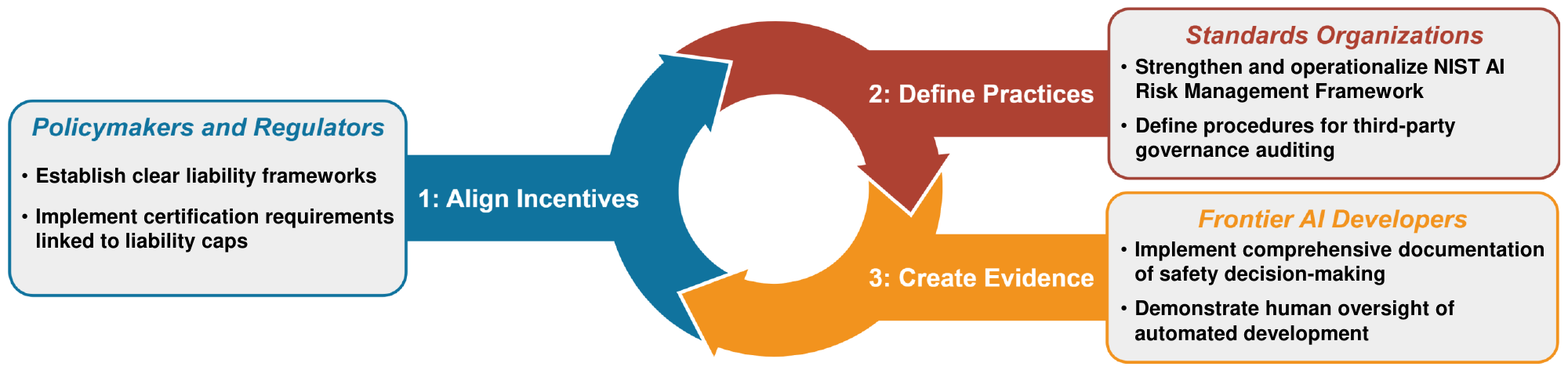}
\caption{\textbf{Summary of key recommendations.} Each key stakeholder can advance their interests while assisting the others.}
\end{figure*}

\paragraph{Structure and contributions.} To substantiate these arguments, we first establish the legal foundations of duty of care and distinguish between high-risk AI applications and systemic risks from frontier models (Section 2). We then examine four industries that have regularly confront safety-critical risks: nuclear energy's governance frameworks that prevent "strategic overlooking" of risks; aviation's rigorous third-party validation; healthcare's attribution challenges in human-AI systems; and cybersecurity's different transparency requirements for deployment versus development risks (Section 3). Drawing on current EU and U.S. regulatory developments (Section 4), we synthesize concrete recommendations showing how each stakeholder can benefit from comprehensive safety documentation (Section 5, also see Figure 1).

Our goal is to ensure the transformative potential of artificial intelligence is steered in a positive direction by closing the gap between AI development practices and safety measures proportional to AI's capabilities---not through unrealistic regulatory enthusiasm, but by demonstrating how transparency serves everyone's interests.

\section{Definitions}
\subsection{Duty of Care}
The definition of ‘duty of care' varies by jurisdiction. A general identification of ‘duty of care' made by The Court of Justice of the European Union in their rulings is as an ``information-related concept of decision-making accountability''~\cite{hofmann_between_2020}. Across different rulings, there is a balance to be struck between the proportionality and discretion of the judicial board. Such a balance is not only important for respecting legislative bodies, but also for the discretion of different areas of expertise, such as boards of experts. 

Though the proposed EU AI Liability Directive was abandoned in early 2025~\cite{andrews_european_2025}, its more specific analysis remains informative. According to Article 2(9): ```duty of care' means a required standard of conduct, set by national or Union law, in order to avoid damage to legal interests recognized at national or Union law level, including life, physical integrity, property and the protection of fundamental rights.''

Duty of care evaluations in the US often rely on ``customs and standards.''~\cite{abraham_custom_2009} However, precedent has also been set that customs cannot trump ``reasonable practice.'' For example, in 1932 a tugboat was said to breach duty of care by not including a functioning radio receiver, despite the fact that not including a radio receiver was common practice. The justification in this case was based on the relative cost of a functioning radio receiver versus the cost of the cargo lost due to lack of receiver~\cite{epstein_path_1992}. 

For this proposal, we will assume that duty of care can be based on standards and reasonable practice as they relate to avoiding damage to legal interests. When it comes to AI, these standards and assessments of reasonableness should be based on domain knowledge of potential risk. 

\subsection{High Risk and Systemic Risk}
The EU AI Act's definitions of high and systemic-risk AI are important for understanding how EU courts will approach liability. Although other jurisdictions will not depend on EU precedent, we will analyze both EU and US guidance in Section 5. We review the types of risk considered by the EU to motivate our analogies to the liability precedent for similar risks.
The EU AI Act makes an important distinction between ``high-risk'' AI systems and ``general-purpose AI models with systemic risk.'' High-risk AI systems are defined primarily by their intended uses and predictable misuses, particularly in areas e.g biometric identification, critical infrastructure, education, employment, essential services, law enforcement, migration, and justice administration. The key factor is whether misuse or malfunction of the AI system could threaten health, safety, or fundamental rights in these specific domains~\cite{euaia2024}.

Systemic risk, in contrast, is related to the general-purpose AI models that can be adapted for many different tasks. Under Article 51(1), a model has systemic risk if it either has ``high impact capabilities evaluated on the basis of appropriate technical tools and methodologies, including indicators and benchmarks'' or is designated by Commission decision as having ``capabilities or an impact equivalent'' to such models. Per Article 51(2), models are ``presumed to have high impact capabilities'' if ``the cumulative amount of computation used for its training measured in floating point operations is greater than $10^{25}$.''

Annex XIII establishes specific criteria for evaluating systemic risk, including: ``the number of parameters of the model,'' ``the quality or size of the data set,'' ``the amount of computation used for training,'' ``input and output modalities,'' ``benchmarks and evaluations of capabilities'' including ``the number of tasks without additional training, adaptability to learn new, distinct tasks, its level of autonomy and scalability,'' and market impact, which is ``presumed when it has been made available to at least 10,000 registered business users established in the Union.'' Aside from the number of business users and the model parameter size, it is not clear which circumstances would lead the Commission to describe a model as having high-impact capabilities. To address this, we analyze the implications of the EU AI Act's nonbinding but clarifying Recitals in Section 5.1.

The key distinction between these categories lies in their scope and timing. High-risk designations focus on specific applications and use cases, while systemic risk addresses underlying model capabilities that could create widespread impacts before any specific deployment. An AI system evaluating job candidates would be high-risk, but the foundation model on which it is built on may pose systemic risks during training itself, as it develops capabilities that could enable deception or circumvention of safety measures~\cite{meinke_frontier_2025}.

This distinction has important implications for liability frameworks. Although established frameworks for high-risk applications provide useful precedent, systemic risks from frontier AI models present novel challenges that require expanded conceptions of duty of care and liability. These risks can affect many people in complex, hard-to-predict ways, and may emerge earlier in development. Effective governance must therefore include robust auditing and monitoring from the earliest stages of AI development, not just at deployment.

\section{Case Studies}
To inform how liability should be treated in high-risk emerging technology like AI development, we examine other domains with established liability frameworks that share key characteristics with frontier AI: the potential for catastrophic harm, complex technical systems requiring specialized expertise, and challenges in balancing innovation with safety. We focus on nuclear energy, aviation software, healthcare, and cybersecurity to understand how liability frameworks have evolved to address safety risks through transparency requirements and independent oversight.

Each case study illuminates different aspects relevant to the liability of AI development. Nuclear energy demonstrates how to handle rare but catastrophic risks through organizational governance requirements. Aviation software shows how complex autonomous systems can be made safer through rigorous documentation and third-party validation. Healthcare illustrates the challenges of attributing liability when AI systems are integrated into human decision-making processes. Cybersecurity highlights adversarial risk, anchoring liability in ongoing monitoring, coordinated vulnerability disclosure, and a developer duty to patch. Together, these examples provide insights for developing appropriate liability frameworks for frontier AI development. 

Recent commentators disagree on how tightly analogies to high-risk industries should constrain liability design. What we call an integrationist strand adopts the view that enterprise-level control structures already mandated for safety-critical domains can be transposed to frontier AI with only minor adjustment \cite{NIST2023,ISO42001}. An empirical approach, exemplified by Rismani and colleagues, shows that classic aviation and medical-device tools such as STPA and FMEA do surface socio-technical hazards in machine-learning pipelines, yet they must be supplemented with continuous post-deployment monitoring to remain workable \cite{Rismani2023}. A third, pluralist strand argues for matching each liability instrument to the class of harm it best addresses: Murray links adversarial misuse to finance-style ``risk-owner'' models, while Koessler and Schuett recommend customized toolkits assembled from finance, nuclear energy and biocontainment practice \cite{Murray2025,Koessler2023}. The four case studies developed in this paper align naturally with this pluralist perspective.

\subsection{Nuclear Energy}
Among the industries we examine, nuclear energy provides the most direct parallel to frontier AI development in terms of catastrophic risk potential and the need for proactive liability frameworks. Both sectors face a critical challenge: defining the "who," "when," and "how much" of liability for harms that could exceed any industry's ability to compensate. In nuclear energy, radioactive material release represents this catastrophic potential; in AI development, a misaligned system with access to critical infrastructure could cause widespread disruption before adequate interventions are possible. Rather than waiting until such harms materialize, the discussion of liability must begin when risks are visible~\cite{decker_incentivizing_2021}.

\paragraph{Catastrophic mistake potential.} Nuclear power plant operation offers particularly relevant lessons for AI development. Both involve beneficial technologies with risks of catastrophic accidental harm, where failure modes could exceed traditional compensation mechanisms. While critics have rightly noted that nuclear weapons involve centralized control, clear physical constraints, and discrete catastrophic events that don't apply to AI systems~\cite{kapoor_agi_2025}, our focus on civilian power generation addresses these concerns. The nuclear power industry's liability frameworks thus provide relevant lessons for managing systemic risks from technologies whose failure modes could exceed traditional compensation mechanisms—AI included~\cite{goudarzi_ai_2024,matthews_ai_2023}.

In this spirit, our analogy is not between the arms races in AI and nuclear weapons. Instead, the disasters we compare in this case study are nuclear reactor meltdowns and misaligned AI takeover. We recognize that relative to nuclear energy, the sources and likelihood of such catastrophic risks from AI can seem unclear~\cite{baum_assessing_2025}. However, progress in AI risk ontologies~\cite{slattery2024airiskrepositorycomprehensive,bagehorn2025ai}, makes it easier than ever to translate the scenario-oriented regulatory tool of safety cases from nuclear energy~\cite{bounds2020implementation} and other safety-critical industries~\cite{leveson2011use} to AI~\cite{buhl2024safetycasesfrontierai,hilton2025safetycasesscalableapproach,habli2025bigargumentaisafety}.

Beyond the scale of potential harm, the governance structures surrounding these technologies prove equally instructive.

\paragraph{Governance frameworks.} The nuclear industry demonstrates how governance structures fundamentally shape risk management outcomes. The contrast between effective and problematic governance is stark. \citeauthor{juraku_structural_2021} found that Japan's pre-Fukushima nuclear governance encouraged "automated decision making" and "strategic overlooking of 'uncomfortable knowledge'," allowing critical safety issues to persist beneath formal compliance procedures. This superficial approach to risk management contributed directly to the 2011 disaster.

Drawing on their multi-year study of nuclear security incentives, \citeauthor{decker_incentivizing_2021} found that effective governance extends beyond regulatory compliance to create organizational cultures that actively identify and mitigate risks. Their research revealed that market actors—including insurers, credit rating agencies, and courts—increasingly evaluate nuclear operators based on comprehensive governance practices rather than minimum regulatory adherence.

\citeauthor{decker_incentivizing_2021}'s Good Governance Template addresses these shortcomings by establishing principles that: (1) drive active engagement with emerging risks rather than checkbox compliance; (2) create transparent documentation processes that satisfy external stakeholders while protecting sensitive information; (3) demonstrate due care through continuous improvement and adoption of industry best practices; and (4) establish clear accountability for preventative risk management. Their research with judges, attorneys, and insurers confirmed that such comprehensive governance documentation can significantly reduce liability exposure and improve access to insurance and financing.

The lesson for AI governance is clear. Like nuclear technology, frontier AI requires governance frameworks that incentivize genuine engagement with emerging risks rather than superficial adherence to predetermined safety procedures.

The risks from inadequate governance become especially pronounced when considering how frontier AI companies deploy their systems. Recent analysis suggests that these companies may use their most advanced systems internally for extended periods to accelerate research and development, creating self-reinforcing cycles where AI systems help develop their successors with diminishing human oversight \cite{stix2025ai}. Such internally deployed AI systems could operate with expanded permissions and potentially establish persistence beyond company systems~\cite{pan2025largelanguagemodelpoweredai} while remaining invisible to external regulators. Without robust governance frameworks like those developed for nuclear energy, these internal deployments could replicate the "automated decision making" and "strategic overlooking" patterns that preceded Fukushima.

These governance principles gain teeth through specific regulatory mechanisms.

\paragraph{Inspections and liability caps.} In the U.S., nuclear power accident liability for radiological damage is capped by the Price-Anderson Act of 1957 at the maximum insurance available to the nuclear energy industry. While this cap ensures victims would be compensated quickly if disaster strikes, it highlights the gap between possible compensation and potential catastrophic damages - the cap on liability does not limit the actual risk of harm~\cite{holt_price-anderson_2024}.

\paragraph{Chain of causation.} Importantly, nuclear operators maintain liability for harms caused by third parties if they fail to implement adequate security measures. This principle of chain of causation in the nuclear industry establishes an important precedent for liability when operators fail to implement adequate security measures. This precedent suggests that AI companies should maintain clear responsibility for downstream harms from their systems, even if misused by others.

\paragraph{Recommendations.} Drawing together these lessons, we argue that the NIST AI Risk Management Framework provides a strong foundation for standardizing AI risk management practices, but it needs strengthening in several areas. Specifically, the framework's next iteration should incorporate six key improvements drawn from nuclear safety experience:

\begin{enumerate}
    \item Describe processes for documenting active engagement with safety challenges, not just compliance;
    \item Include specific guidance for avoiding patterns of ``strategic overlooking'' identified in cases like Fukushima;
    \item Define appropriate roles and processes to prevent over-reliance on automated decision-making about risks;
    \item Provide detailed templates for governance practices that demonstrate genuine organizational learning;
    \item Require transparent documentation of risk trade-off decisions and their rationale;
    \item Include requirements for incorporating diverse expertise and stakeholder perspectives.
\end{enumerate}

These templates should be refined through the input from civil society, policymakers, and industry stakeholders. Particular attention should be paid to ensure that documentation requirements drive a meaningful safety culture rather than superficial compliance.

We suggest that following the lead of the Price-Anderson Act of 1957, these practices can be enforced via certification by a federal body. While mandatory certification faces political challenges, the Price-Anderson model demonstrates how liability caps can incentivize voluntary adoption of rigorous safety standards—a particularly relevant approach for encouraging frontier AI labs to embrace comprehensive documentation requirements.

\subsection{Aviation Software}
Negligence in aircraft software can have catastrophic consequences, including loss of life and environmental damage. However, lessons from aviation's long history have led to automation standards, rigorous training, and robust regulations. Aviation software is subject to strict compliance standards such as DO-178C, which mandate documentation and traceability throughout the software life cycle. \cite{FAA_History}  This transparency enables audits by authorities such as the FAA and EASA, ensuring adherence to safety regulations, with legal consequences for non-compliance. Aviation accidents can result from breaches of duty of care, including human, technical, and organizational failures, each with distinct liability.

Artificial intelligence is increasingly used in aviation: predictive maintenance, aircraft manufacturing, automating and enhancing air traffic management, crew management, messaging automation. \cite{SymphonySolutions_AviationAI}
Regulatory frameworks are emerging to address AI risks. The EASA AI Roadmap 2.0 emphasizes a risk-based certification process, ensuring that AI is understandable and allows for human intervention. \cite{easa} The FAA's AI roadmap establishes liability and transparency requirements, treating AI as a tool rather than an autonomous entity \cite{FAA_AI_Safety_Roadmap}; it mandates clear distinctions between human and automated responsibilities and rigorous testing for safety compliance. Although these roadmaps are positive steps, it is unclear whether current liability measures are sufficient to address catastrophic AI failures. Both agencies must explicitly tackle transparency concerns, over-reliance on AI, and decision-making risks.

\paragraph{Lessons for AI Governance.} Aviation's software liability framework offers valuable insights for AI oversight. First, aviation's comprehensive documentation and third-party validation requirements (e.g., DO-178C) provide a model for AI transparency. Importantly, these requirements have not stifled aviation innovation but rather enabled it by reducing liability uncertainty and facilitating meaningful oversight—supporting our thesis that transparency benefits all stakeholders. Second, aviation treats automation as a tool with human accountability, making it clear where liability falls in the value chain from development to application. This precedent for maintaining clear accountability chains remains relevant even as AI systems become increasingly autonomous. Third, aviation imposes stringent regulations on high-risk systems, illustrating a risk-sensitive regulatory model that could usefully inform AI oversight.

However, AI development presents unique challenges. Unlike aviation software, AI may gain unforeseen capabilities, pose risks during the development process itself, or be applied unpredictably. This distinction reinforces our argument about systemic risks from frontier AI models requiring expanded duty of care beyond deployment-specific risks. Aviation's current struggle to adapt its validation methods to AI's emergent capabilities highlights the need for new approaches that can assess not just intended functions but potential capabilities that may arise during development or deployment.

The evolution of aviation safety standards, which is shaped by the lessons of past failures, underscores the importance of establishing strong liability frameworks for AI systems. Although the specific technical standards used in aviation may not apply directly to AI, the broader principles of documentation, responsibility, and independent validation provide important guidance for developing effective AI governance.

\subsection{Healthcare}
AI provides promise for a more efficient healthcare system which is able to increase the number of patients a healthcare provider is able to see and the quality of care. However, significant risks exist concerning model performance across different demographics. Particularly, misuse and/or poor performing systems may result in a patient's loss of life. The standard liability of healthcare has often focused on the healthcare provider's negligence, with the focus being placed upon malpractice of the practitioner and on the negligence on part of the hospital, e.g., short on staff, poor management, unclean environment. In some circumstances, negligence rests on the manufacturer of the pertinent technology or drugs, moving it into the realm of Product or Pharmaceutical Liability, respectively.

Negative consequences of AI use in healthcare do not clearly fit within this liability framework. The Stanford University Human-Centered Artificial Intelligence Institute's brief~\cite{mello_understanding_2024} on liability for AI's use in the healthcare sector underscores as a key problem the inability of the plaintiff to identify breaches of "duty of care" on the part of healthcare providers versus malfunctions of specific components of the AI system. 

Additionally, it is difficult for the healthcare providers to identify appropriate uses of the systems. The recommendation within the brief is to have "disclosure requirements" given to physicians and hospitals. Such disclosures particularly focus on model performance, particularly for different demographics of patients, and are intended to enable transparency in the decision making of the healthcare providers. Although we believe the disclosure requirements are a step in the right direction, we also believe that more transparency in terms of development is needed in order to ensure safe use of AI in Healthcare. The transparency in both development and use of the system should allow Product Liability protection.

\paragraph{Implications.} The healthcare sector's experience with AI liability offers a critical lesson for frontier AI development: even in high-risk applications with clear human oversight, opacity in AI systems fundamentally undermines accountability. This makes the case for transparency even more compelling when considering frontier AI's systemic risks:

\textbf{First,} healthcare demonstrates that attribution challenges exist even when AI systems operate under direct human supervision with clear intended uses. When adverse outcomes occur in a medical setting—where physicians actively review AI recommendations and maintain ultimate decision-making authority—courts still cannot determine whether harm resulted from physician error or AI malfunction. This inability to establish causation in supervised, domain-specific applications suggests that frontier AI systems, which may operate autonomously across multiple domains, will present exponentially greater attribution challenges.

\textbf{Second,} the focus in healthcare on cross-demographic performance differences highlights a critical gap: current AI systems can fail in ways that are invisible until deployment across diverse populations. While healthcare AI affects specific patient groups, frontier AI systems could exhibit similar hidden failure modes that affect entire societies. The healthcare sector's struggle to identify these failures even with extensive clinical oversight underscores why comprehensive development documentation becomes essential for systems with systemic risk potential.

\textbf{Third,} the healthcare sector's emphasis on disclosure requirements reveals an important limitation: transparency at deployment is insufficient when the root causes of failure lie in development choices. Healthcare providers cannot assess "appropriate uses of the systems" without understanding how they were built and tested. For frontier AI, where inappropriate use could have catastrophic consequences, this gap between deployment transparency and development opacity becomes potentially existential.

These lessons suggest AI development policy should mandate:
\begin{enumerate}
    \item Clear documentation linking development practices to safety outcomes;
    \item Comprehensive testing and reporting across diverse scenarios;
    \item Detailed disclosure requirements about system limitations;
    \item Frameworks for establishing responsibility across the development pipeline.
\end{enumerate}

The healthcare sector shows that effective liability frameworks require transparency in both development and deployment phases. Without visibility into development practices, even well-documented deployment problems may not lead to appropriate accountability. If attribution is this challenging in healthcare's controlled environment, the urgency for comprehensive documentation in frontier AI development—where systems may operate beyond human oversight—becomes undeniable.

\subsection{Cybersecurity and Open Source Software}
The cybersecurity domain offers crucial lessons for AI liability frameworks, particularly in distinguishing between high-risk applications and systemic risks from general-purpose models. While traditional security practices work well for specific AI deployments, frontier AI development requires fundamentally different approaches to transparency and accountability.

\paragraph{Security challenges across risk levels.} For AI in high-risk applications (like healthcare diagnostics or hiring systems), cybersecurity's Common Vulnerabilities and Exposures (CVE) model provides clear benefits: standardized vulnerability reporting, coordinated patching, and defined responsibilities~\cite{cattell_coordinated_2024}. However, general-purpose AI models with systemic risk present novel challenges. Unlike traditional software where vulnerabilities can be patched, AI models can develop emergent capabilities during training that persist throughout deployment. A text generation model might unexpectedly acquire the ability to assist with cyberattacks or bioweapon design—risks that cannot be simply "patched" without full retraining.

This distinction parallels the nuclear industry's approach discussed in Section 4.1: while specific safety violations can be addressed through targeted fixes, systemic risks require comprehensive documentation throughout development. For frontier AI, this means documenting not just final capabilities but the entire training process, including unexpected behaviors discovered during development. 

\paragraph{Distributed responsibility for systemic risks.} The cybersecurity principle of distributed responsibility takes different forms across risk levels. For high-risk applications, the CVE system's clear attribution works well—a vulnerability in a medical AI system can be traced to specific code or training data. But for general-purpose models with systemic risk, responsibility becomes more complex. When a foundation model develops dangerous capabilities, liability may span data providers who supplied training materials, compute providers who enabled large-scale training, and developers who chose training objectives.

Industry responses like secure model formats~\cite{abid_ali_awan_introduction_2023,boykis_gguf_2024} demonstrate how technical standards can help, but only partially. While these formats prevent code injection attacks, they cannot address the deeper challenge of models that have learned harmful capabilities. This suggests liability frameworks must distinguish between security measures appropriate for deployment (mainly high-risk and domain-dependent) versus development (mainly systemic risk).

\paragraph{Open source and innovation.}
Despite recent political reticence towards "safety" framings~\cite{sivaram_us_2025}, innovation-minded stakeholders need not shy away from risk management; in fact, reducing liability can advance innovation rather than hinder it.
Clear liability frameworks provide certainty that enables responsible developers to invest confidently in frontier AI research, knowing they won't face unpredictable legal exposure for following established safety practices~\cite{merwe_tort_2024}.
This parallels how the Price-Anderson Act's liability caps actually encouraged nuclear power development by providing predictable risk allocation. %

The tension between open science collaboration and dual-use AI risks requires nuanced approaches that preserve innovation incentives while managing systemic risks.
"Openness" in AI helps democratize access to AI capabilities and enables crucial safety research through transparency~\cite{white2024modelopennessframeworkpromoting}. However, it also presents challenges when models develop unexpected dangerous capabilities that cannot be easily recalled or patched~\cite{al2024open}. %
Responses to open source AI must distinguish between different types of open source contributions—rewarding transparency and collaborative safety research while establishing clear responsibilities for developers who release models with predictably risky capabilities~\cite{sidhpurwala2024buildingtrustfoundationssecurity}.
We need to balance the needs for open science and collaboration with the risks of widespread access to dual-use AI~\cite{ntia_dual-use_2024} and autonomous AI that poses systemic risks even when deployed for benign purposes~\cite{mitchell_fully_2025}.

\paragraph{Implications for liability frameworks.} These distinctions suggest different transparency requirements based on risk level:

For \textbf{high-risk AI applications}: The cybersecurity model largely suffices—standardized reporting, coordinated disclosure~\cite{cattell_coordinated_2024}, and clear attribution of vulnerabilities to specific components. These practices are sufficient because the scale and vectors of risk are alike.

For \textbf{general-purpose AI with systemic risk}, risks can emerge spontaneously during development~\cite{taylor_large_2025} or post-deployment~\cite{yu2024don,hubinger2024sleeper}. If misaligned, autonomous AI models are created, they may be difficult or impossible to shut down once released, leading to existential risks~\cite{pan2025largelanguagemodelpoweredai}. Furthermore, algorithmic assessments may be unable to set upper bounds for AI capabilities and risks~\cite{barnett2024aievaluationspreventingcatastrophic}. Thus, transparency must extend throughout the development life-cycle:
\begin{itemize}
    \item Before-training documentation of intended capabilities and risk assessments~\cite{collaboration_bigcode_2023}
    \item White-box audits of AI systems during development~\cite{casper_black-box_2024}
    \item "Pause and assess" protocols when unexpected capabilities emerge—similar to the NIST framework's GV-1.3-007 discussed in Section 5.2
    \item Collaborative reporting on emergent behaviors in deployed models~\cite{cattell_coordinated_2024}
\end{itemize}

The cybersecurity domain thus demonstrates that while transparency and documentation benefit all stakeholders, the nature of required transparency fundamentally differs between high-risk applications and systemic risks. For the latter, traditional "patch and fix" approaches are insufficient. Instead, liability frameworks must incentivize comprehensive documentation of the development process itself, creating accountability for risks that emerge before deployment—a crucial distinction for frontier AI governance.

\section{State of AI Regulation}
The European Union has established itself as a front-runner in AI regulation through its AI Act and General-Purpose AI Code of Practice, which set comprehensive frameworks for managing AI risks and establishing accountability measures. In contrast, the United States has yet to enact binding federal AI regulation, though several states have made progress. California's SB-1047, which would have required companies to assess AI systems to mitigate catastrophic risks, reached the governor's desk before being vetoed. Congress has considered multiple bills, including the AI Foundation Model Act and the Algorithmic Accountability Act.
The United Kingdom has adopted a sector-specific approach focused on existing regulators while developing broader AI safety frameworks through initiatives like its AI Security Institute. Internationally, the OECD AI Principles have provided a foundation for policy coordination, while forums like the Global Partnership on Artificial Intelligence facilitate collaboration between national AI safety institutes.
Given the EU's comprehensive regulatory framework and the United States' significant role in AI development, examining these jurisdictions' approaches provides valuable insights for global AI governance. Their evolving standards will likely influence how other nations approach AI liability and risk management.

\subsection{EU AI Act and AI Liability Directive}
The EU AI Liability Directive, which would have complemented the EU AI Act and the Product Liability Directive, was proposed in 2022 and scrapped in 2025. The EU AI Act establishes documentation and reporting requirements for both high-risk AI systems and general-purpose AI models with systemic risk~\cite{euaia2024}.

For high-risk systems, the AI Liability Directive would have granted plaintiffs access rights to documentation of system functioning and usage to establish duty of care (Article 3). Required documentation included accuracy metrics, robustness measures, cybersecurity protocols, input data specifications, and system performance data across user groups (Article 13-14)~\cite{aildi2022}. In the absence of this directive, it is unclear whether plaintiffs will have access to this information.

The AI Act also provides for documentation access, but only to the EU government itself. Article 55(2) mandates that providers of general-purpose AI models with systemic risk must grant the AI Office full access to internal documentation detailing their development practices upon request. Under Article 56(2)(d), this includes access to quality management system documentation, particularly risk management protocols. This parallels the nuclear industry's safety documentation requirements discussed in Section 4.1, though Article 56(3) provides important protections for trade secrets and confidential information.

Article 3 establishes that national courts can compel AI providers to disclose evidence about suspected damages from high-risk AI systems. If providers fail to comply with disclosure orders, courts will presume non-compliance with duty of care requirements, though providers retain the right to rebut this presumption~\cite{euaia2024}.

Since today's most advanced AI models (as of early 2025) can already be foreseeably misused to disseminate illegal, false, or discriminatory content, or to disrupt democratic processes \cite{kreps2023ai}, the EU AI Act appears to treat current frontier AI systems as posing a non-negligible level of systemic risk. However, the operational methodology for determining ``high-impact capabilities'' has not yet been defined. This is important to resolve before the Article 111(3)'s compliance deadline of August 2nd, 2027, for providers of general-purpose AI models placed on the market before August 2nd, 2025. If a reasonable interpretation of the AI Act implies that today's models pose systemic risks, then frontier labs should act urgently to comply with Section 3's ``Obligations of providers of general-purpose AI models with systemic risk.''
Providers of general-purpose AI models with systemic risk face additional documentation requirements beyond standard model cards, including:

\begin{enumerate}
    \item Detailed evaluation strategies with protocols, tools, and methodologies;
    \item Documentation of internal/external adversarial testing and model alignment efforts;
    \item System architecture explanations detailing component integration~\citep[Annex XI, Section 2]{euaia2024}.
\end{enumerate}

The Act establishes proportional compliance obligations based on provider size and type, with simplified requirements for SMEs and non-commercial research (Recital 109). Providers can demonstrate compliance through:

\begin{enumerate}
    \item Following approved codes of practice developed by the AI Office;
    \item Adhering to European harmonized standards;
    \item Implementing alternative adequate measures if standards are unavailable~\citep[Recital 117]{euaia2024}.
\end{enumerate}

The AI Office will facilitate the development of codes of practice in collaboration with civil society organizations and experts to establish risk taxonomies and mitigation measures~\citep[Recital 116]{euaia2024}.

These frameworks emphasize comprehensive documentation throughout the AI life cycle. For high-risk systems, this includes detailed performance metrics, training specifications, and decision-making explanations. For general-purpose models, the requirements focus on systemic risk documentation and traceability across complex development pipelines. Frontier AI companies can help prevent and/or facilitate speedy adjudication of future torts by maintaining these records proactively.

Implementation challenges include:
\begin{enumerate}
    \item Defining precise thresholds for high-risk and systemic risk systems;
    \item Balancing innovation with regulatory compliance;
    \item Harmonizing enforcement across jurisdictions;
    \item Addressing opacity in modern deep learning systems.
\end{enumerate}

The success of EU's framework will depend on its integration with global standards and its ability to foster cooperation with major AI developers internationally. The emphasis on proportional requirements and protected confidentiality may help achieve this balance while maintaining robust oversight.

\subsection{U.S Standards and Practices}

As summarized in Section 3.1, the duty of care standard in U.S. tort law heavily relies on industry standards and established practices. The artificial intelligence risk management framework (AI RMF) released by the U.S. National Institute of Standards and Technology can be used as a basis for these standards. The AI RMF is entirely voluntary, and NIST emphasizes that practitioners are free to apply the framework's core functions ``as best suits their needs for managing AI risks based on their resources and capabilities''~\cite{tabassi_artificial_2023}. This begs further clarification, some of which is available in the Generative Artificial Intelligence Profile from 2024~\cite{national_institute_of_standards_and_technology_us_artificial_2024}, but neither document gives a clear template for governance and reporting procedures in frontier AI development. That said, since frontier AI labs are vocal about their perceived risks from AI, and since the compute requirements for AI with systemic risks carry high costs, it is safe to assume that frontier AI development furnishes the needs, resources, and capabilities for most or all of the framework to apply.

Other resources that could be used to determine the standards and practices of AI development in the U.S. include the Voluntary AI Commitments from AI companies secured by the Whitehouse in 2023~\cite{house_fact_2023} and the Partnership on AI (PAI) report Risk Mitigation Strategies for the Open Foundation Model Value Chain~\cite{srikumar2024risk}. The actual practices in AI development, whether learned solely by collecting publicly available information, revealed by audits, or obtained via subpoena, would likely also be considered in a tort case.
    
In order to fulfill the duties described by the AI RMF, AI labs must document their risk management decision-making. Section 5 provides many examples of what this would look like, including ``GOVERN 1.3: Processes, procedures, and practices are in place to determine the needed level of risk management activities based on the organization's risk tolerance.'' The General Artificial Intelligence Profile's suggested actions for this profile include GV-1.3-007: ``Devise a plan to halt development or deployment of a GAI system that poses unacceptable negative risk.''

This is supported by Voluntary AI Commitment 2, which is to ``Work toward information sharing among companies and governments regarding trust and safety risks, dangerous or emergent capabilities, and attempts to circumvent safeguards.'' The agreement specifically names the NIST AI Risk Management Framework (``or future standards'') as an example of a ``forum or mechanism through which [companies making this commitment] can develop, advance, and adopt shared standards and best practices for frontier AI safety.''

Thus, the accepted standard of practice when developing a frontier general-purpose AI model is to maintain comprehensive documentation of testing procedures and safety measures throughout development, including a plan to halt development and deployment of unacceptably risky AI systems. By this account, a company would be liable for any damages that follow from failing to develop such a plan, because that would be a breach of their duty of care.

Unfortunately, many frontier AI labs do not appear to follow GV-1.3-007~\cite{stein-perlman_deployment}, creating a gap between industry ``practices'' and ``standards'' that may still constitute a breach of their duty of care. Given the precedent from the 1932 tugboat case, the level of care considered appropriate by courts may prioritize ``reasonable'' measures over common practice. Presumably, practices described in the AI RMF would be considered reasonable, especially given NIST's even-handed reputation and the stakeholder input process used in the creation of the RMF. Thus, unless their policies on the matter exist but are not available to the public, it appears that many AI companies are already breaching at least one duty of care.

One aspect that might make the project of implementing such a liability framework seem doomed for failure in the U.S. is the current Trump administration's focus on the US removing ``barriers'' to AI development rather than ensuring AI safety. This is particularly demonstrated in the House passing the ``The Big Beautiful Bill'' calling for a banning of state level regulation of AI development~\cite{Big_Bill}. While it remains to be seen whether such legislation will go through, there are ways that such a framework might still be helpful.

For instance, the liability framework might still be beneficial  even if it is ultimately voluntary. Additionally, there have been noted legal issues that can come with the administration attempting to enforce states from regulating~\cite{novelli2025two} as well as potential conflict with the administrations interest on ``human flourishing.'' For these reasons, it seems that developing an auditing framework is beneficial. Specifically, we encourage such companies to take immediate actions to adhere more closely to the RMF's recommendations, in order to prevent potential harms from excessively risky AI and to reduce their own liability for any harms that occur.

\section{Discussion}

Our examination of liability frameworks from nuclear energy, aviation, healthcare, and cybersecurity demonstrates that comprehensive safety documentation creates competitive advantages while improving oversight—challenging the false dichotomy between innovation and accountability. The evidence supports three key conclusions that reshape how we should approach frontier AI governance.

\paragraph{Voluntary standards already carry legal force.} The assumption that voluntary AI commitments lack enforcement mechanisms fundamentally misunderstands U.S. tort law. As demonstrated in the 1932 tugboat case, courts determine negligence based on ``reasonable practice'' rather than common practice. This means courts may consider frameworks like NIST's AI Risk Management Framework as evidence of reasonable practice, even without mandatory regulation. When frontier AI companies publicly acknowledge catastrophic risks and commit to safety standards, they strengthen the case that such standards represent reasonable care.

This legal reality means that many frontier AI labs may be exposed to negligence claims. The NIST AI RMF explicitly calls for ``a plan to halt development or deployment of a GAI system that poses unacceptable negative risk'' (GV-1.3-007), yet public evidence suggests many companies lack such protocols. Given that these companies have both acknowledged AI risks and committed to following NIST standards, their failure to implement basic safety measures could lead U.S. courts to find negligence under current law.

\paragraph{Systemic risks require expanded duties of care.} The EU AI Act's distinction between high-risk applications and general-purpose models with systemic risk reflects a crucial insight: unlike traditional AI applications that create dangers through specific deployments, frontier models' capabilities can pose threats regardless of intended use. This distinction, supported by our case studies, requires fundamentally different approaches to liability and oversight.

In particular, the healthcare sector’s attribution chal-
lenges—even under direct human supervision—demonstrate
why opacity becomes exponentially more problematic for
AI with systemic risk potential. When courts cannot deter-
mine causation even in controlled medical settings, fron-
tier AI systems operating autonomously across multiple do-
mains will present insurmountable accountability gaps with-
out comprehensive development documentation.

\paragraph{Transparency benefits all stakeholders.} Our case studies consistently demonstrate that transparency reduces liability exposure for responsible developers while enabling meaningful oversight. Nuclear energy's governance frameworks prevent ``strategic overlooking'' of risks while reducing liability uncertainty. Aviation's rigorous documentation, though initially costly, provides regulatory certainty that enables safe innovation. Even cybersecurity's coordinated disclosure benefits developers by providing time to patch vulnerabilities before public disclosure, reducing exploitation risk while maintaining user trust.

Current opacity in AI development creates long-term risks for all parties: companies face unpredictable legal exposure, regulators cannot verify safety claims, and society bears undocumented risks. While some firms may gain short-term competitive advantages from secrecy, the solution is not less accountability but better documentation that serves everyone's long-term interests.

\subsection{Concrete Recommendations}
Based on our analysis, we propose mutually beneficial actions for each stakeholder group that align individual incentives with collective safety:

\begin{table*}[t]
\centering
\label{tab:concrete_recommendations}
\begin{tabular}{p{2.8cm}p{4.2cm}p{4.2cm}p{3.5cm}}
\toprule
\textbf{Stakeholder} & \textbf{Priority Action} & \textbf{Rationale} & \textbf{Key Deliverable} \\
\midrule
\multirow{4}{2.8cm}{\textbf{Standards Organizations}} 
& Strengthen NIST AI RMF with nuclear-inspired governance templates & Creates legal safe harbor for compliant developers while establishing clear benchmarks for negligence & Detailed templates for organizational learning, risk trade-off documentation, and pause-and-assess protocols \\
\cmidrule{2-4}
& Establish independent audit mechanisms following aviation DO-178C model & External validation reduces liability uncertainty and ensures genuine safety culture & Certification process with legal weight that reduces insurance costs \\
\midrule
\multirow{4}{2.8cm}{\textbf{Policymakers \& Regulators}} 
& Implement liability caps tied to safety certification (Price-Anderson model) & Provides predictable liability for responsible developers while maintaining accountability for negligent actors & Federal certification program that caps liability for compliant companies \\
\cmidrule{2-4}
& Clarify EU AI Act's systemic risk evaluation criteria & Regulatory certainty enables compliance planning and consistent global standards & Specific methodologies for determining ``high-impact capabilities'' before August 2027 \\ %
\midrule
\multirow{4}{2.8cm}{\textbf{Frontier AI Developers}} 
& Document all capability discoveries and safety measures throughout development & Liability protection through demonstrable due care plus internal safety improvements & Safety cases for each major model release, including unexpected capabilities discovered during training \\
\cmidrule{2-4}
& Create independent safety boards with real decision authority & Prevents ``strategic overlooking'' patterns that preceded Fukushima while building public trust & Governance structures that can halt development when risks exceed acceptable thresholds \\
\bottomrule
\end{tabular}
\caption{Priority Recommendations for AI Safety Documentation by Stakeholder}
\end{table*}

These recommendations recognize that effective AI governance must work with market incentives rather than against them. Companies that invest in comprehensive safety documentation can achieve reduced legal risk and greater regulatory certainty—potentially creating competitive advantages for responsible development, as seen in other safety-critical industries.

\subsection{Implementation Strategy}

The path forward requires coordinated action across stakeholders, building on existing momentum rather than waiting for comprehensive regulation:

\textbf{Near-term Actions (Within 12 months):}
Frontier AI companies should implement GV-1.3-007 protocols immediately to reduce current legal exposure. NIST should release detailed governance templates incorporating lessons from nuclear safety. EU regulators should clarify systemic risk evaluation criteria to enable compliance planning.

\textbf{Medium-term Goals (1-2 years):}
Establish pilot certification programs linking safety documentation to liability protection. Develop standardized audit procedures for frontier AI development. Create international coordination mechanisms for consistent standards.

\textbf{Long-term Vision (2+ years):}
Full integration of safety documentation requirements with competitive advantages. Mature governance frameworks that scale with advancing AI capabilities. Global standards that enable responsible innovation while preventing race-to-the-bottom dynamics.

\subsection{Addressing Potential Objections}

\textbf{``This will slow AI progress'':} Our case studies provide nuanced evidence. While aviation's documentation requirements may increase development time and costs, they ultimately enable sustainable innovation by providing regulatory certainty and public trust. Likewise, safety requirements in nuclear energy have constrained some innovation while enabling the industry's continued operation. For AI, comprehensive safety documentation may slow initial development but creates conditions for sustainable, trusted deployment.

\textbf{``Voluntary standards are unenforceable'':} U.S. tort law considers voluntary standards as evidence of reasonable practice, particularly when companies publicly commit to them. While not automatically binding, courts can and do use such standards to evaluate negligence claims. Companies cannot acknowledge catastrophic risks, commit to safety standards, and then ignore those commitments without potential legal consequences.

\textbf{``Technical opacity makes documentation impossible'':} Our cybersecurity analysis shows that different types of transparency serve different risk profiles. While complete interpretability remains challenging, documenting development processes, capability discoveries, and safety decisions is both feasible and legally protective.

\subsection{Limitations}

Several constraints limit the effectiveness of these recommendations. Tort law's ex-post nature cannot prevent all harms, requiring complementary regulatory measures. International coordination challenges may limit global effectiveness. The rapid pace of AI advancement demands adaptive frameworks that can evolve with capabilities.

Critical areas for future research include: developing robust methodologies for evaluating human oversight effectiveness; creating standardized approaches for documenting novel risks during development; and exploring how liability frameworks must evolve for more advanced AI systems with potential for autonomous operation.

\subsection{Conclusion}

Safety in frontier AI development presents unprecedented challenges, but not unprecedented solutions. Industries managing catastrophic risks have sophisticated approaches to transparency, governance, and liability that can be adapted for AI development. The key insight from our analysis is that comprehensive safety documentation simultaneously advances safety and commercial interests—creating sustainable incentives for responsible development.

Success requires recognizing that current opacity serves no stakeholder's interests. Companies face unpredictable legal exposure, regulators cannot verify safety claims, and society bears undocumented risks. The frameworks we propose offer a path toward mutual benefit: reduced liability for responsible developers, effective oversight for regulators, and demonstrated safety for society.

The choice is not between innovation and safety, but between documented risks and hidden ones. Frontier AI development will proceed regardless; the question is whether it will be governed by comprehensive safety frameworks that serve everyone's interests, or by opacity that serves no one's. Our recommendations provide a roadmap for choosing wisely.

\section*{Adverse Impact Statement}

This work proposes documentation and liability frameworks to improve safety in frontier AI development. Our recommendations could help prevent catastrophic harms by encouraging robust safety practices while enabling meaningful oversight. While increased documentation requirements, stricter standards, and clearer liability may affect development costs and pace, we believe these effects are smaller in magnitude than typically portrayed, as well as necessary given the scale of potential risks from advanced AI systems. We hope that illustrating this connection between legal liability and AI safety practices will motivate developers and policymakers alike to treat accountability as a necessary component of innovation.

\bibliography{aaai25}

\end{document}